# Gas Phase Chemistry of Cool Exoplanet Atmospheres: Insight from Laboratory Simulations

Chao He[1], Sarah M. Hörst[1], Nikole K. Lewis[2], Julianne I. Moses[3], Eliza M.-R. Kempton[4], Mark S. Marley[5], Caroline V. Morley[6], Jeff A. Valenti[7], & Véronique Vuitton[8]

[1] Department of Earth and Planetary Sciences, Johns Hopkins University, Baltimore, MD, USA che13@jhu.edu

[2]Department of Astronomy and Carl Sagan Institute, Cornell University, 122 Sciences Drive, 14853, Ithaca, NY, USA

[3]Space Science Institute, Boulder, CO, USA

[4]University of Maryland, College Park, MD, USA

[5]NASA Ames Research Center, Mountain View, CA, USA

[6]The University of Texas at Austin, Austin, TX, USA

[7]Space Telescope Science Institute, Baltimore, MD, USA

[8]Université Grenoble Alpes, Grenoble, FR

## Abstract:

Photochemistry induced by stellar UV flux should produce haze particles in exoplanet atmospheres. Recent observations indicate that haze and/or cloud layers exist in the atmospheres of exoplanets. However, photochemical processes in exoplanetary atmospheres remain largely unknown. We performed laboratory experiments with the PHAZER chamber to simulate haze formation in a range of exoplanet atmospheres (hydrogen-rich, water-rich, and carbon dioxide-rich at 300, 400, and 600 K), and observed the gas phase compositional change (the destruction of the initial gas and the formation of new gas species) during these experiments with mass spectrometer. The mass spectra reveal that distinct chemical processes happen in the experiments as a function of different initial gas mixture and different energy sources (plasma or UV photons). We find that organic gas products and $O_2$ are photochemically generated in the experiments, demonstrating that photochemical production is one of the abiotic sources for these potential biosignatures. Multiple simulated atmospheres produce organics and $O_2$ simultaneously, which suggests that even the co-presence of organics and $O_2$ could be false positive biosignature. From the gas phase composition changes, we identify potential precursors ($C_2H_2$, HCN, $CH_2NH$, HCHO, etc.) for haze formation, among which complex reactions can take place and produce larger molecules. Our laboratory results indicate that complex atmospheric photochemistry can happen in diverse exoplanet atmospheres and lead to the formation of new gas products and haze particles, including compounds ($O_2$ and organics) that could be falsely identified as biosignatures.

**Abstract Graphic:**

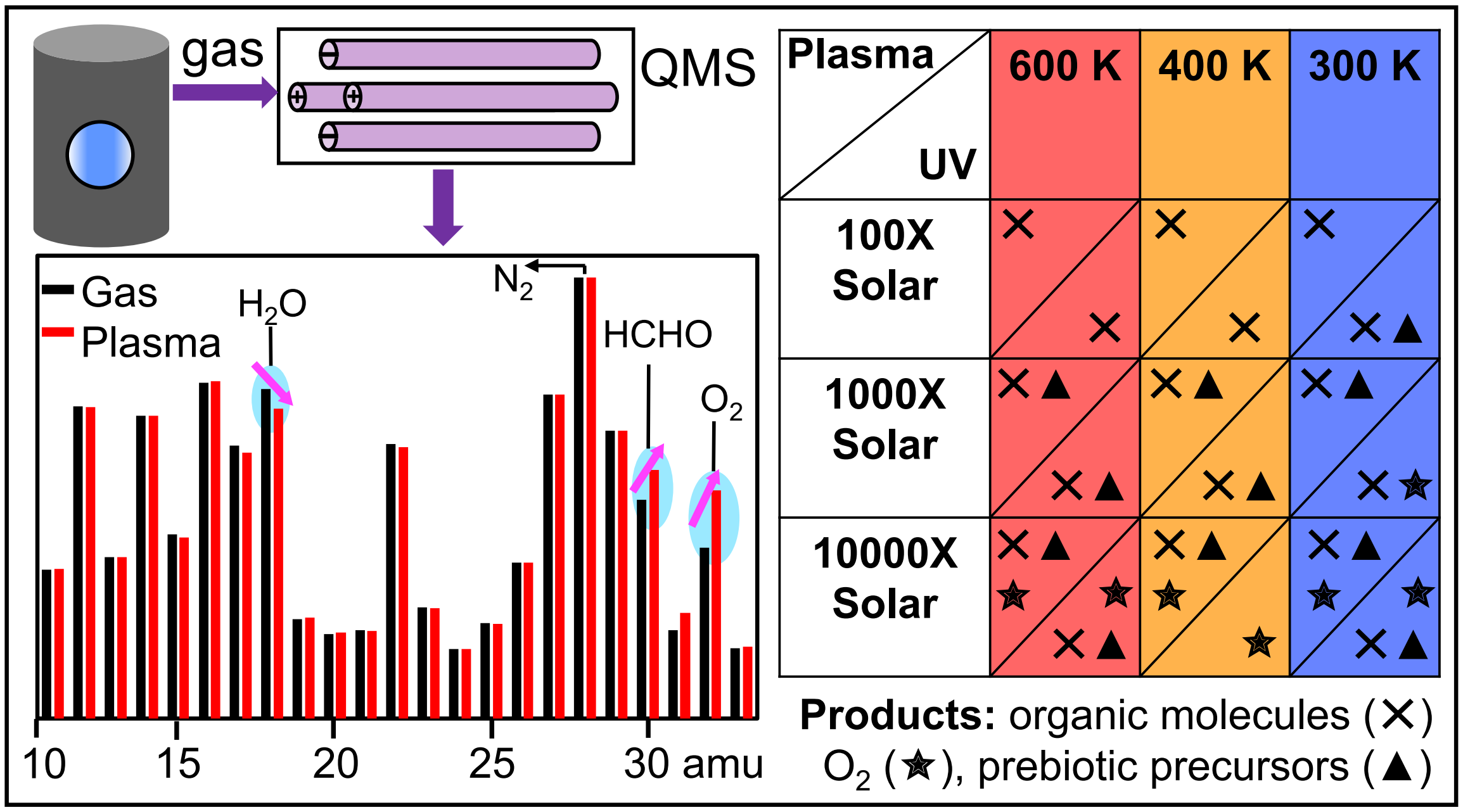

**Keywords: Exoplanet, atmosphere, photochemistry, biosignature, prebiotic chemistry, organic haze**

## 1. INTRODUCTION

Thousands of exoplanets have been discovered in the last two decades, but the majority of them do not have an analogue in our solar system, such as those planets with size or mass between Earth's and Neptune's (i.e., super-Earths and mini-Neptunes). These two types of exoplanets are the most abundant types of planets and are expected to exhibit a wide variety of atmospheric compositions[1-5]. The Transiting Exoplanet Survey Satellite (TESS) mission, which launched in April 2018, will find more super-Earths and mini-Neptunes for atmospheric characterization by the James Webb Space Telescope (JWST) and other future telescopes. Clouds and/or hazes are present in every solar system planetary atmosphere and are expected to be present in exoplanet atmospheres based on our understanding of particle formation in planetary atmospheres. Recent observations have shown that clouds and/or hazes play a significant role in the atmospheres of small, cool exoplanets[6-10], such as GJ 1214b, HD 97658b, GJ 436b, and GJ 3470b. Although a variety of equilibrium cloud decks are expected at various temperatures as specific molecular and atomic species condense in the atmosphere, photochemical hazes can potentially be formed over a range of temperatures, pressures, and atmospheric compositions[11-14].

However, very little is known about the photochemical processes for haze formation in these exoplanet atmospheres as the atmospheric phase space (sub-Neptune atmospheres in the 300–600 K temperature range) has been largely unexplored. This lack of knowledge about haze formation will limit our ability to interpret the exoplanet observations and characterize exoplanet atmospheres. It is extremely challenging to theoretically simulate the complex chemical processes for haze formation over a broad range of exoplanet atmospheres. Laboratory simulations have improved our understanding of haze formation in Solar System bodies (e.g., Titan[15]) and could provide critical information on haze formation/properties for exoplanets. We have performed a series of laboratory atmosphere simulation experiments[13,14,16] that explored a broad range of atmospheric parameters relevant to super-Earths and mini-Neptunes. We reported haze production rates and the particle size distributions measured in these experiments[13,14,16]. These experiments started with nine different gas mixtures that yielded a wide variety of haze particle production rates. We monitored the gas composition with a mass spectrometer during these experiments. Here we present the gas phase chemistry in the experiments. We first evaluate the destruction of the original gases and identify newly formed gas species from the mass spectra, and then try to find out the gas precursors that may be indicative of photochemistry and haze formation. We also investigate connections between gas composition and haze production rate and explore possible photochemical pathways in these atmospheric scenarios.

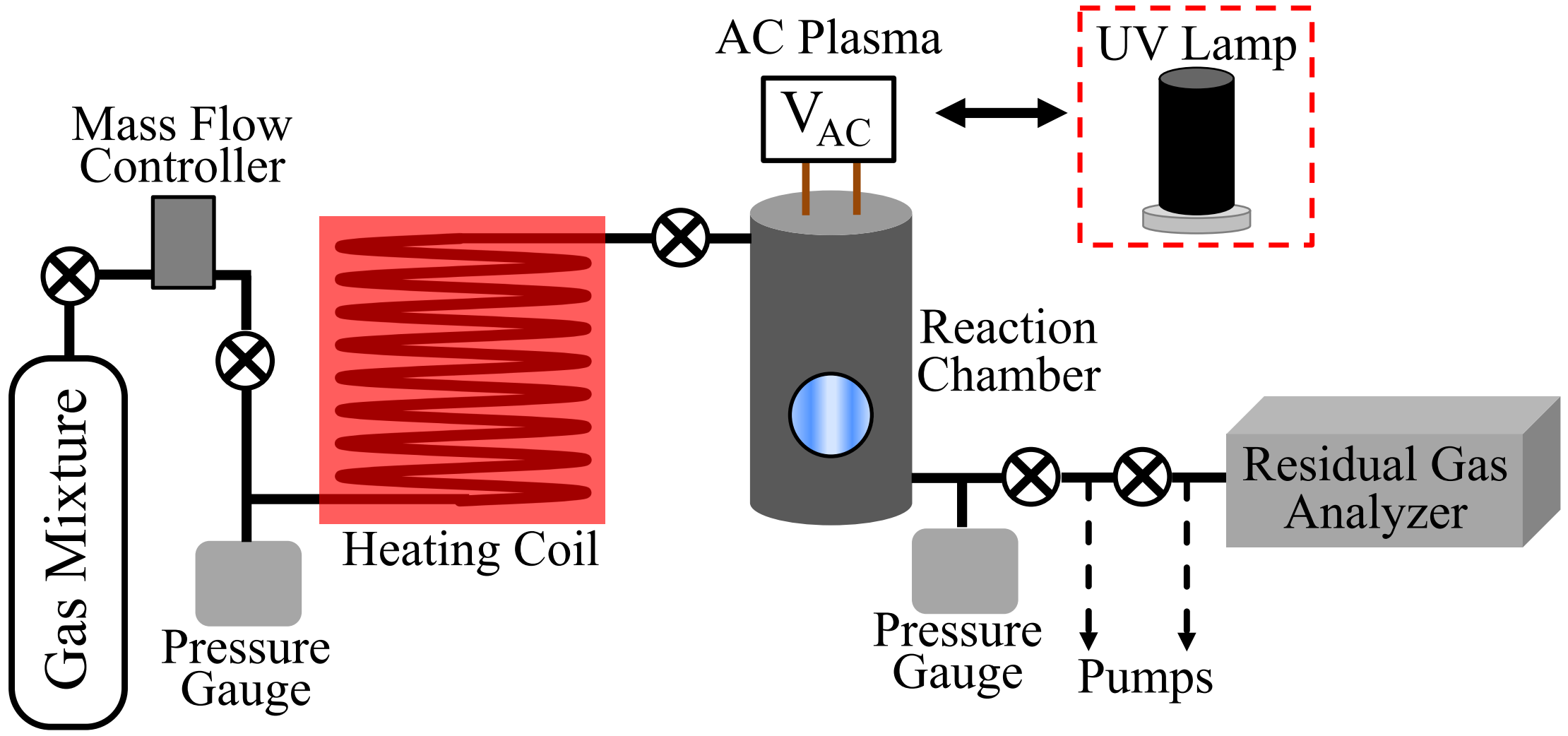

Figure 1. Schematic of the PHAZER experimental setup used for this work. The PHAZER setup allows us to use one of two energy sources: a cold plasma generated by an AC glow discharge or FUV photons produced by a hydrogen lamp. Note that the schematic shown here provides a concept of our setup. The details of the setup varied due to the large variety of gases used for these experiments. Detailed description of the setup and experimental procedure can be found in our previous papers[13,14,16].

## 2. MATERIALS AND EXPERIMENTAL METHODS

### *2.1. Haze Production Setup*

Figure 1 shows a schematic of the Planetary Haze Research (PHAZER) experimental setup at Johns Hopkins University[13,17]. The PHAZER chamber allows us to conduct simulation experiments over a broad range of atmospheric parameters with one of two energy sources (AC plasma or FUV photons). Figure 2 shows the initial gas mixtures for our experiments, calculated from the chemical equilibrium models[3] for 100×, 1000×, and 10000× solar metallicity over a range of temperatures (300, 400, and 600 K). More details about the chemical equilibrium model and the gas mixtures can be found in previous studies[3,13,14,16].

Metallicity is a simple proxy that captures the enhancement of elements heavier than hydrogen over solar nebular values. The higher metallicity represents more heavier molecules (hydrogen-poor) in the atmosphere. A much more compositionally diverse range of atmospheres are expected for hydrogen-poor atmospheres, which could lead to a much wider variety of atmospheric chemistries. A deeper investigation into gas-phase chemistry and haze production in hydrogen-poor atmospheres is warranted, and for the time being, the 10000× solar metallicity atmosphere is representative of one possible outcome for such an atmosphere. Many exoplanet atmosphere forward models and retrieval codes assume thermochemical equilibrium, so it is important to investigate experimentally how photochemistry might alter the equilibrium gas concentrations in an exoplanet atmosphere.

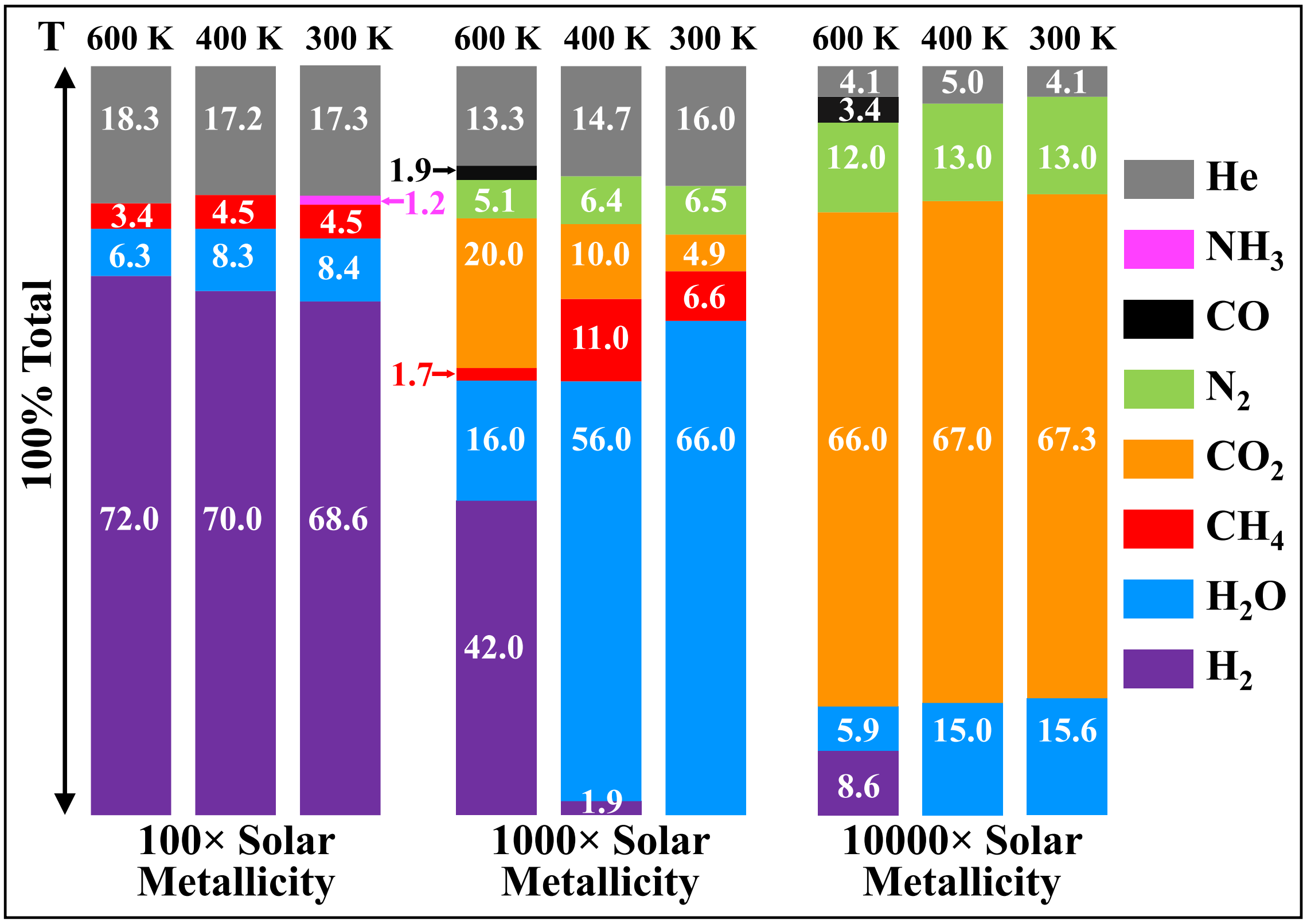

Figure 2. The initial gas mixture compositions used in our experiments[13,14,16]. The phase space spans temperature from 300 to 600 K and metallicity from 100 to 10,000× solar. Note that equilibrium chemistry calculations assuming various enhancements over solar nebular values were used to determine initial gas mixtures[3]. The 100× ($H_2$-rich), 1000× ($H_2$-rich for 600 K, or $H_2O$-rich for 400 and 300 K), and 10000× ($CO_2$-rich) solar metallicity here is used to describe the abundance of heavier molecules present in exoplanet atmospheres. We only included gases with a calculated abundance of ≥1% to maintain a manageable level of experimental complexity. The pressure, temperature and gas compositions used in the experiments are self-consistent based on the model calculations.

As discussed previously[13,14,16], the chemical equilibrium calculated atmospheric composition scenarios provide a reasonable starting point for our study. Premixed gas mixture flows through a heating coil, which heats the gas mixture to the experimental temperature (600, 400, or 300 K). The gas flow rate is 10 sccm (standard cubic centimeters per minute) and the pressure in the reaction chamber is maintained at a few mbar (depending on the temperature). The heated gas mixture is exposed to AC glow discharge (plasma), or UV photons from a UV lamp (HHeLM-L from Resonance Ltd., a hydrogen light source producing UV radiation from 110 to 400 nm). The plasma or the UV radiation dissociates and/or ionizes the reactant gases and initiates chemical reactions in the chamber. Newly formed gas products and remaining reactant gases flow out of the chamber, while any solid produced in the experiment remains. The experiments run continuously for 72 hours and gases flowing out of the chamber are monitored with a Residual Gas Analyzer (RGA, a quadrupole mass spectrometer). The detailed experimental procedure was described previously[13,14,16].

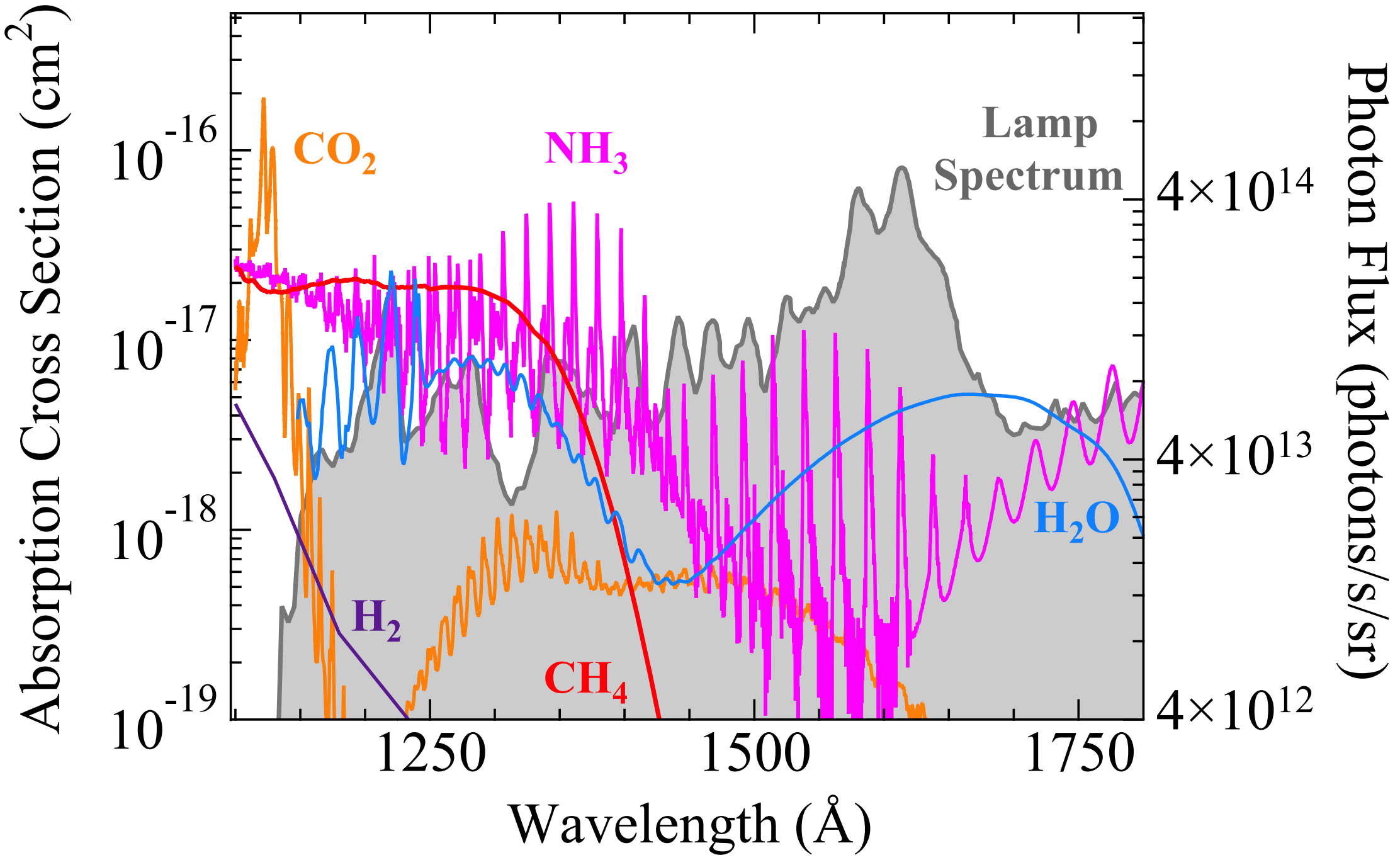

Figure 3. The spectrum of the UV lamp used in these experiments (right axis) (as provided by the manufacturer, Resonance Ltd.) and the absorption cross sections of the gases we used in these experiments (left axis). Note that the cross sections of $N_2$ and CO are not shown here because they are very low (near zero) in this wavelength range.

UV radiation and electrical discharge are the two main types of energy sources used to mimic one or several of the energy sources capable of generating radicals and other activated species in planetary atmospheres[15]. Since it is impossible to perfectly replicate the energy environment of a planetary atmosphere in the lab, we compare the results from two different energy sources. UV photons are the main drivers of photochemistry in atmospheres of solar system bodies (Venus, Earth, Jupiter, Saturn, Titan, and Pluto), and stellar UV radiation should induce photochemistry in exoplanet atmospheres. The UV lamp we used here is similar to that used for simulating photochemistry in the atmospheres of early Earth and Titan[18-22]. The spectrum (110 to 180 nm) of the lamp and the absorption cross sections of the gases used in our experiments are shown in Figure 3. The photons produced by the lamp (110 to 400 nm) are not sufficiently energetic to directly dissociate very stable molecules such as $N_2$ or CO, but previous studies with similar UV lamps suggested that nitrogen can be incorporated into the molecular structure of haze particles through an unknown photochemical process[19,23].

The energy inputs of charged particles (i.e., cosmic rays, magnetospheric protons, and electrons) could also play a significant role in the upper atmosphere. The plasma produced by the AC glow discharge might be able to simulate electrical activities and/or charged particles in planetary upper atmospheres. The plasma can provide sufficient energy to directly break very stable molecules such as $N_2$ or CO. In addition, the AC glow discharge is a cold plasma source and does not alter the neutral gas temperature significantly, which is often used as an analogue for the relatively energetic environment of planetary upper atmospheres[15].

The AC glow discharge and the UV photons we used in the experiments are two different types of energy sources to simulate different processes in planetary atmospheres. Because photochemistry in planetary atmospheres usually occurs on extended time scales, laboratory simulations usually use power densities that are greater than what is actually experienced in a real atmosphere[15] in order to observe obvious compositional change and/or yield a sufficient quantity of material for analysis in a reasonable time period. We estimated the energy densities of two different energy sources and compared them to small (2 Earth radius), cool (equilibrium temperature: 300 to 600 K) exoplanets around a given host M-star (3000 K). The energy density of the AC glow discharge is about 170 $W/m^2$, which is ~5 times greater than that of the UV lamp (110 nm to 400 nm, 36 $W/m^2$). The UV flux (110 to 400 nm) from the host M-star that reaches to the exoplanet is about 0.312 $W/m^2$ (300 K equilibrium temperature) to 4.62 $W/m^2$ (600 K equilibrium temperature). The energy densities of both energy sources are higher than that in the hypothetical small, cool exoplanets. In our experiments, 72-hour of exposure to the AC glow discharge roughly corresponds to between 110 days (for 600 K exoplanet) and 1600 days (for a 300 K exoplanet) of UV irradiation from the host M-star; 72-hour of exposure to the UV photons is equivalent to between 23 days (for 600 K exoplanet) and 346 days (for a 300 K exoplanet) of UV irradiation from the given host M-star.

*2.2. Gas Phase Composition Measurement*

A small portion of the gas mixture flowing out of the chamber is channeled into the RGA (Stanford Research Systems) where the gas phase products are analyzed. The RGA is a quadrupole mass spectrometer (QMS) with a mass range of 1 to 300 amu, a resolution of approximately 0.5 amu, and a minimum detectable partial pressure of $10^{-11}$ torr. It has an electron ionization (EI) source and a standard 70 eV energy is used in our measurements. The scanning mass range is 1–100 amu and each scan takes about 2 minutes. Before introducing gas mixture to the RGA chamber, the background in the RGA chamber (a few $10^{-7}$ Torr) is first measured (50 scan). After the background scan, the gas mixture is introduced into RGA chamber (a few $10^{-5}$ Torr) and a 50-scan average mass spectrum of the gas mixture is collected before turning on the plasma or the UV lamp. After turning on the plasma or the UV lamp, we allow the gas mixture to flow for 30 minutes to reach steady state before we take gas phase measurements. In each experiment, we take the RGA measurements multiple times, and the resulted mass spectra are consistent aside from the noise. We average all the scans (1000 scans) for the duration of each experiment to lower the noise level, and obtain an average mass spectrum (MS). The RGA background is removed from both mass spectra of the gas mixture with the plasma (or the UV lamp) off and on. The total intensity of all peaks (1 to 100 amu) in each spectrum is used as a fixed reference to normalize the peaks because the flow rates of the mixtures are the same for all the experiments. Thus, we can compare the normalized mass spectra and find out the changed peaks.

To minimize potential contamination, our chamber was cleaned thoroughly with detergent and solvents, further cleaned by using an ultrasonic cleaner for 30 minutes, and then baked at 400 K for 12 hours. The chamber was pumped down to $10^{-3}$ Torr and was continuously pumped for 24 hours before running the experiment. We also ran a reference experiment without plasma or UV exposure at 600 K with the 600 K-10000× metallicity gas mixture for 72 hours. During the reference experiment, we monitored the gas mixture with the RGA and did not observe any compositional changes in the gas mixture. When analyzing the RGA data from the plasma and UV experiments, we removed the RGA background, normalized the mass spectra, and then determined the changed peaks in the normalized mass spectra. Therefore, the changed peaks are due to the photochemistry induced by plasma or UV photons, rather than contamination from the chamber or the initial gas mixture.

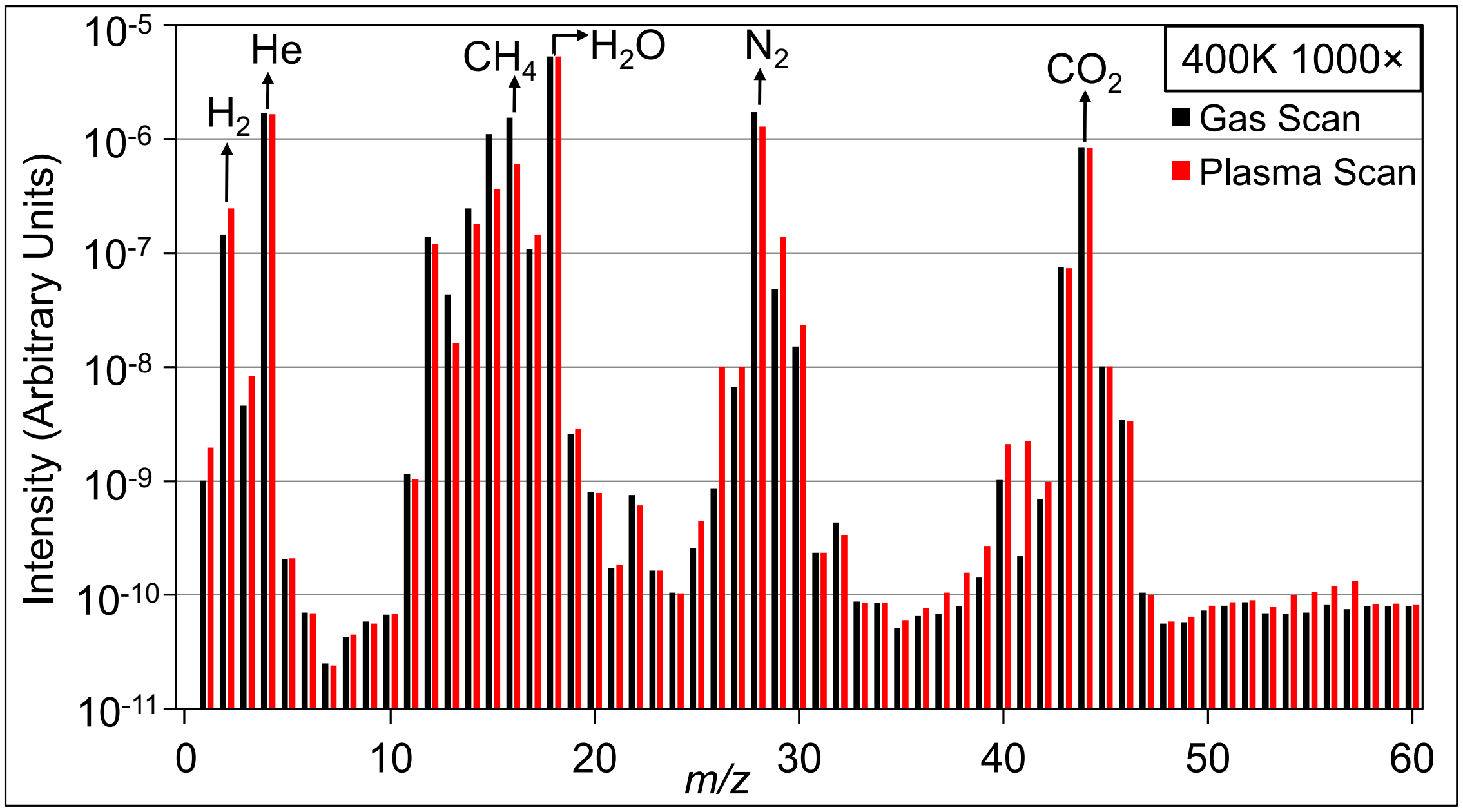

Figure 4. The mass spectra of gas mixture for the 400 K-1000× metallicity plasma experiment, with plasma off (gas scan, black) and on (plasma scan, red). MS peaks from 1 to 60 amu are shown since the heavier peaks are near the noise level ($8\times10^{-11}$). The initial gases, $H_2$, (2 amu, 1.9%), He (4 amu, 14.7%), $CH_4$ (16 amu, 11%), $H_2O$ (18 amu, 56%), $N_2$ (28 amu, 6.4%), and $CO_2$ (44 amu, 10%), are labeled near the molecular peaks. With the plasma on, $CH_4$ and $N_2$ in the initial gas mixture decrease, while $H_2$ increases. New species are generated in the gas phase, including $NH_3$, C2 ($C_2H_2$, HCN, $CH_2NH$, $C_2H_6$/HCHO), C3 [$C_3H_4$ (40), $CH_3CN$ (41), $C_3H_6/CH_2N_2/C_2H_2O$ (42)], and C4 [$C_4H_6/C_2H_2N_2/C_3H_2O$ (54); $C_3H_5N/C_2HON$ (55); $C_4H_8/C_2H_4N_2/C_3H_4O$ (56); $C_3H_7N/CH_3N_3/C_2H_3ON$ (57)] species. It should be noted that C2, C3, and C4 species stand for organic molecules that have 2, 3, and 4 heavy atoms (C, N, and O), respectively.

## 3. RESULTS AND DISCUSSION

### *3.1. Mass Spectra of Gas Phase Products*

Figure 4 shows the mass spectra (MS) of the gas mixture for the 400 K-1000× metallicity plasma experiment with plasma off and on (the RGA background is removed from both spectra). The initial gases are shown in the spectra, including $H_2$, (2 amu), He (4 amu), $CH_4$ (16 amu), $H_2O$ (18 amu), $N_2$ (28 amu), and $CO_2$ (44 amu). Besides the molecular ion peaks, the isotope peaks and fragment peaks (caused by 70 eV electron impact) are also shown in the MS. The species listed in Table 1 are the most probable, most abundant species. The resolution of the mass spectra is not high enough to resolve the species with identical nominal mass for higher mass, thus there are several possibilities for higher mass.

Since we are trying to understand the gas chemistry induced by the energy source, we focus on the peaks that have significant intensity change after turning on the plasma in 400 K-1000× metallicity plasma experiment (Figure 4). Changes in peaks are the result of destruction of the initial gases and/or production of new gas phase species. For instance, the increase at 1, 2, and 3 amu are caused by the production of $H_2$ when the plasma is on. The decrease at 12, 13, 14, 15, and 16 amu are associated with the depleting of $CH_4$, and the decrease at 28 amu is due to $N_2$ depletion. With the plasma on, some $CH_4$ and $N_2$ molecules are dissociated and converted to other species. The increased peaks on Figure 4 indicate the new formed species, which are also listed in Table 1. New gas products include ammonia ($NH_3$, 17 amu), acetylene ($C_2H_2$, 26 amu), hydrogen cyanide (HCN, 27 amu), $CH_2NH$ (29 amu, probably methanimine), $C_2H_6$ (ethane, 30 amu) and/or HCHO (formaldehyde, 30 amu), C3 species, and C4 species. It should be noted that C3 and C4 species stand for organic molecules that have 3 and 4 heavy atoms (C, N, and O), respectively. The increased peak at 40 amu is derived from $C_3H_4$ (the structure could be propyne, cyclopropene, and/or allene), 41 amu is derived from acetonitrile ($CH_3CN$), while 42 amu could be from $C_3H_6$ (propene or cyclopropane), $NH_2CN$ (cyanamide), and/or $CH_2CO$ (ketene). The increased peaks at 54, 55, 56, and 57 amu correspond to newly formed C4 species. It could be $C_4H_6$, $C_2H_2N_2$, and/or $C_3H_2O$ for the peak at 54 amu; $C_3H_5N$ and/or $C_2HON$ for 55 amu; $C_4H_8$, $C_2H_4N_2$, and/or $C_3H_4O$ for 56 amu; $C_3H_7N$, $CH_3N_3$ and/or $C_2H_3ON$ for 57 amu. Each molecular formula can have several different structures (isomers); for example, $C_3H_7N$ has 9 stable isomers. With current information, it is difficult to identify the molecular structures of these species. However, the increased peaks clearly demonstrate the formation of C4 species in the gas phase of the 400 K-1000× metallicity plasma experiment.

Table 1. Assignments of peaks in the mass spectra of the 400 K-1000× metallicity plasma experiment (Figure 4)

| Peak (m/z) | Change | Species |
|---|---|---|
| 1 | Increased | $H^+$: fragment of $H_2$ |
| 2 | Increased | $H_2^+$: molecular peak of $H_2$ |
| 3 | Increased | $HD^+$: isotopic peak of $H_2$; or $H_3^+$: protonated $H_2$ |
| 12 | Decreased | $C^+$: fragment of $CH_4$ and/or other carbon species |
| 13 | Decreased | $CH^+$: fragment of $CH_4$ and/or other carbon species |
| 14 | Decreased | $CH_2^+$: fragment of $CH_4$ and/or other carbon species; or $N^+$: fragment of $N_2$ |
| 15 | Decreased | $CH_3^+$: fragment of $CH_4$ |
| 16 | Decreased | molecular peak of $CH_4$; or $O^+$: fragment of $H_2O$ |
| 17 | Increased | molecular peak of $NH_3$; or $OH^+$: fragment of $H_2O$ |
| 18 | No change | molecular peak of $H_2O$ |
| 25 | Increased | $C_2H^+$: fragment of $C_2H_2$ |
| 26 | Increased | molecular peak of $C_2H_2$ |
| 27 | Increased | molecular peak of HCN |
| 28 | Decreased | molecular peak of $N_2$/$C_2H_4$ |
| 29 | Increased | molecular peak of $CH_2NH$ |
| 30 | Increased | molecular peak of $C_2H_6$/HCHO |
| 38 | Increased | $C_3H_2^+$: fragment of $C_3H_4$ |
| 39 | Increased | $C_3H_3^+$ or $HC_2N^+$: fragment of $C_3H_4$ or $CH_3CN$ |
| 40 | Increased | molecular peak of $C_3H_4$ |
| 41 | Increased | molecular peak of $CH_3CN$ |
| 42 | Increased | molecular peak of $C_3H_6$/$CH_2N_2$/$C_2H_2O$ |
| 44 | No change | molecular peak of $CO_2$ |
| 45 | No change | $^{13}CO_2^+$: isotopic peak of $CO_2$ |
| 46 | No change | $CO^{18}O^+$: isotopic peak of $CO_2$ |
| 54 | Increased | molecular peak of $C_4H_6$/$C_2H_2N_2$/$C_3H_2O$ |
| 55 | Increased | molecular peak of $C_3H_5N$/$C_2HON$ |
| 56 | Increased | molecular peak of $C_4H_8$/$C_2H_4N_2$/$C_3H_4O$ |
| 57 | Increased | molecular peak of $C_3H_7N$/$CH_3N_3$/$C_2H_3ON$/$C_2H_3ON$ |

### *3.2. Gas Molecule Changes during the 100× metallicity Experiments*

In order to clearly show the gas phase composition change, we subtract the mass spectrum of the initial gas mixture from the gas mass spectrum with the plasma (or the UV lamp) on, and find the peaks with significant change. For those changed peaks that correspond to gas molecules, we plotted them against m/z, as shown in Figures 5, 7, and 9. Those changed peaks that are mostly due to fragments (like $CH_4$ fragments) are not shown in the figures. It should be noted that 3 spectra (600, 400, and 300 K) are shown in the same plot to provide more information at once but not to compare the intensity ratios from one experiment to another since the initial mixtures are different and the changes in chemical pathways are not only dependent on temperature or energy source. In these figures, the negative peaks represent the gas molecules whose abundances decrease, while the positive ones are the gas molecules whose abundances increase. After turning on the plasma (or the UV lamp), some gas molecules are dissociated and converted to other species, thus leading to the decrease of some compositions in the initial mixture and the increase of newly formed products.

Table 2. The initial gas compositions for the 100× solar metallicity experiments.

| 600 K | 400 K | 300 K |
|---|---|---|
| 72% $H_2$ | 70% $H_2$ | 68.6% $H_2$ |
| 6.3% $H_2O$ | 8.3% $H_2O$ | 8.4% $H_2O$ |
| 3.4% $CH_4$ | 4.5% $CH_4$ | 4.5% $CH_4$ |
| 18.3% He | 17.2% He | 1.2% $NH_3$ |
| | | 17.3% He |

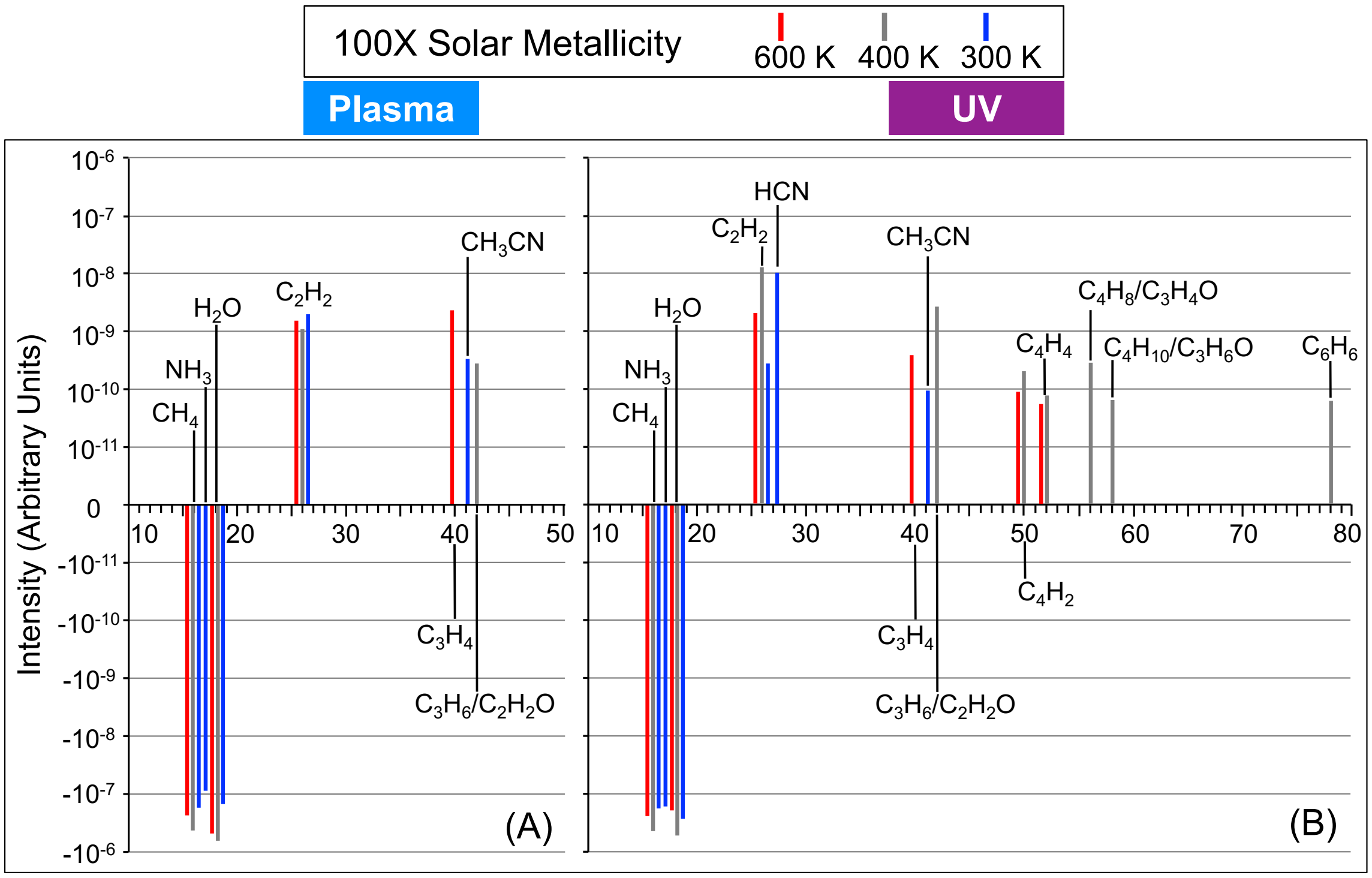

Figure 5. The changed gas peaks in the 100× metallicity experiments for both energy sources: plasma (A) and UV (B). Results are shown in different colors for different temperatures: red, 600 K; orange, 400 K; blue, 300 K. The chemical formulas are labeled near the peaks (Peaks are offset from the nominal mass to show three spectra in one plot).

The initial gas compositions for the 100× metallicity experiments are shown in Table 2. Figure 5 shows the changed gas peaks in the 100× metallicity experiments for both energy sources: plasma (A) and UV (B). The 100× metallicity experiments started with $H_2$-dominated gas mixtures for all three temperatures. In both plasma and UV experiments, the abundances of $H_2$ and He do not change significantly, but the other components in the initial gas mixture all decrease. As shown in Figure 5, $CH_4$ (16) and $H_2O$ (18) decrease for all three temperatures, and $NH_3$ (17) (that is only present in the 300 K case) also decreases. However, the newly formed products vary with the energy sources and temperature. For the plasma experiments, acetylene ($C_2H_2$) and $C_3H_4$ (propyne most likely) are formed in the 600 K experiment; $C_2H_2$ and $C_3H_6$ (propene or cyclopropane) are formed in the 400 K case; $C_2H_2$ and acetonitrile ($CH_3CN$) are formed in the 300 K case. For the UV experiments, $C_2H_2$, $C_3H_4$ (propyne most likely), 1,3-butadiyne ($C_4H_2$), and $C_4H_4$ (1-buten-3-yne or cyclobutadiene) are formed in the 600 K experiment; $C_2H_2$, $C_3H_6$ (propene or cyclopropane), four different 4-carbon species ($C_4H_2$, $C_4H_4$ $C_4H_8$, and $C_4H_{10}$) and $C_6H_6$ (benzene) are formed in the 400 K case; $C_2H_2$, hydrogen cyanide (HCN) and $CH_3CN$ are formed in the 300 K case. For the UV experiment at 400 K, mass 42, 56 and 58 could also be oxygen-containing species ($C_2H_2O$, $C_3H_4O$, and $C_3H_6O$, respectively) since $H_2O$ loss is observed during the reaction. It is important to note that $C_2H_2$ is produced in all the 100× metallicity experiments, suggesting that it may play a significant role in these experiments.

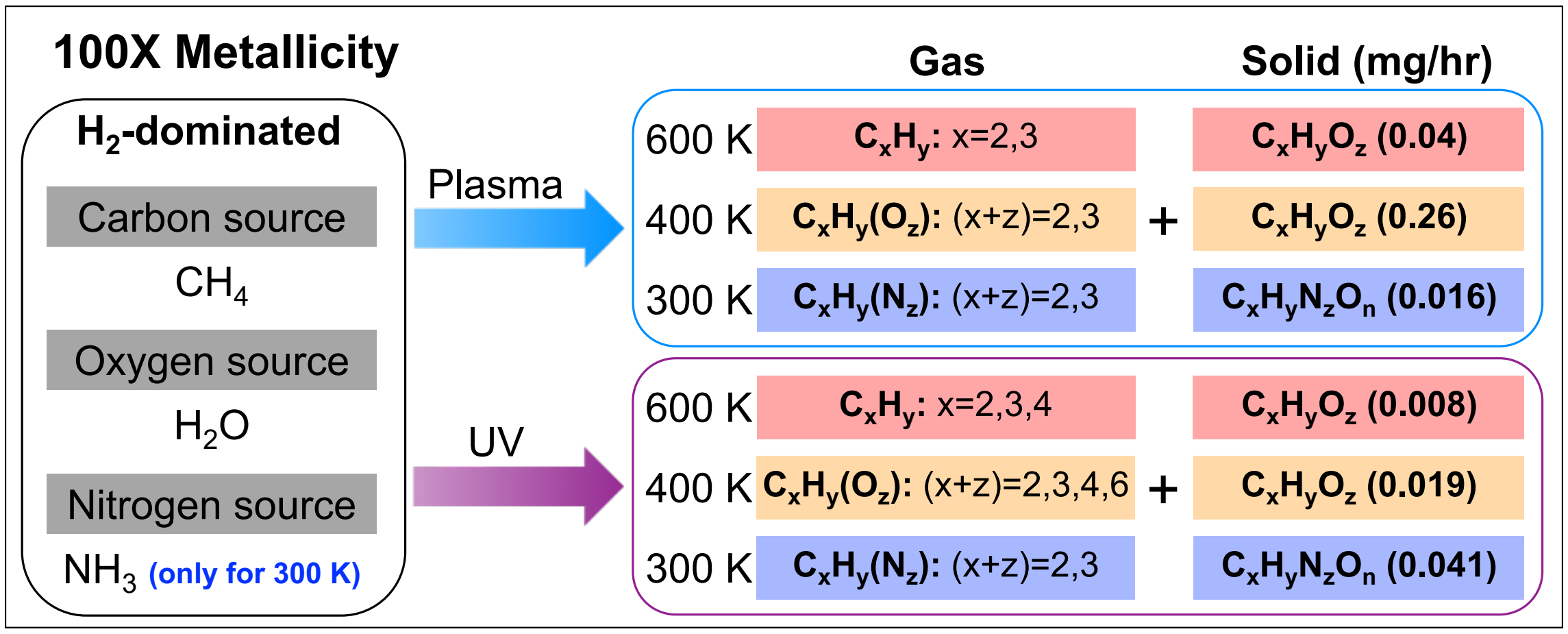

Figure 6. End products of the 100× metallicity experiments. The chemical formula with parentheses represents multiple forms; for example, $C_xH_y(O_z)$ includes $C_xH_y$ and $C_xH_yO_z$. The number in the right column is the haze production rate in each experiment.

The gas composition changes indicate that different chemical pathways resulted in gas to particle conversion in these experiments. The energy input, plasma or UV radiation, modifies the initial gases and starts reactions in the system, eventually resulting in the creation of new gas and solid phase species. Figure 6 shows end products of the 100× metallicity experiments. The initial gas mixtures for the 100× metallicity experiments are $H_2$-dominated at all three temperatures. From the gas phase mass spectra, we learned that the destruction of $CH_4$ and $H_2O$ provides carbon and oxygen sources for both energy inputs (plasma and UV) at three temperatures, and additionally ammonia provides a nitrogen source for the 300 K experiments. Under plasma or UV radiation, these initial gases undergo complex chemical processes and produce new gas and solid phase compounds. The new gas phase products are hydrocarbons, oxygen-containing, or nitrogen-containing species, up to C6 species (Figure 5 and 6). The processes lead to the formation of different gas products for different cases, but also affect the formation and growth of the solid particles. From the destroyed and new species in the gas phase, we can infer the elemental compositions of the solid species. For both the plasma and UV experiments, the solid samples should consist of carbon (C), hydrogen (H), and oxygen (O) at 600 K and 400 K; those at 300 K should also have nitrogen (N). The elemental composition of the solids will be confirmed by further composition analysis, such as high-resolution mass spectrometry.

The gas phase chemical processes also determine the production rate of the solids. The production rates in all the 100× metallicity experiments are relatively low for both energy sources. As previously reported, the presence of $H_2$ in the initial gas mixture reduces haze particle formation[24]. The $H_2$-dominated gas mixture in our experiments provides a highly reducing environment, which could inhibit chain growing and thus decrease the formation of larger molecules. However, both the plasma and the UV photons initiated complex reactions in such reducing environments, which produced new gas products and solid particles. For example, hydrocarbons up to C6 species (benzene) are detected in the gas phase of the 100× metallicity experiment with UV at 400 K. $C_2H_2$ is formed in all 100× metallicity experiments, which could be a key precursor for heavier molecules. $C_2H_2$ is readily formed from photochemistry of $CH_4$ as previously reported in laboratory simulations of Titan's atmosphere[25,26], and possible formation mechanism has been proposed in photochemical models[27]. $C_2H_2$ can react with a wide range of compounds (including itself) to generate bigger molecules[27,28]. In our 100× metallicity experiments, newly formed $C_2H_2$ can serve as a precursor and react with other species in the system to form more complex compounds in both gas phase and solid phase.

*3.3. Gas Molecule Changes during the 1000× metallicity Experiments*

Table 3. The initial gas compositions for the 1000× solar metallicity experiments.

| 600 K | 400 K | 300 K |
|---|---|---|
| 42% $H_2$ | 56% $H_2O$ | 66% $H_2O$ |
| 20% $CO_2$ | 11% $CH_4$ | 6.6% $CH_4$ |
| 16% $H_2O$ | 10% $CO_2$ | 6.5% $N_2$ |
| 5.1% $N_2$ | 6.4% $N_2$ | 4.9% $CO_2$ |
| 1.9% CO | 1.9% $H_2$ | 16% He |
| 1.7% $CH_4$ | 14.7% He | |
| 13.3% He | | |

The initial gas compositions for the 1000× metallicity experiments are shown in Table 3. Figure 7 shows the changed gas peaks in the 1000× metallicity experiments. The 600 K-1000× experiment started with a $H_2$-dominated gas mixture. For the 600 K experiments with both energy sources, we did not observe significant changes for $H_2$, He, or $CO_2$, but saw decreases for $CH_4$ (16), $H_2O$ (18), and $CO/N_2$ (28). At 600 K, $C_2H_2$, HCN, and $CH_3CN$ are formed in the plasma experiment; $C_2H_2$, HCN, $CH_2NH$, and $C_2H_7N/HCONH_2$ (43) are generated in the UV experiment (Figure 7).

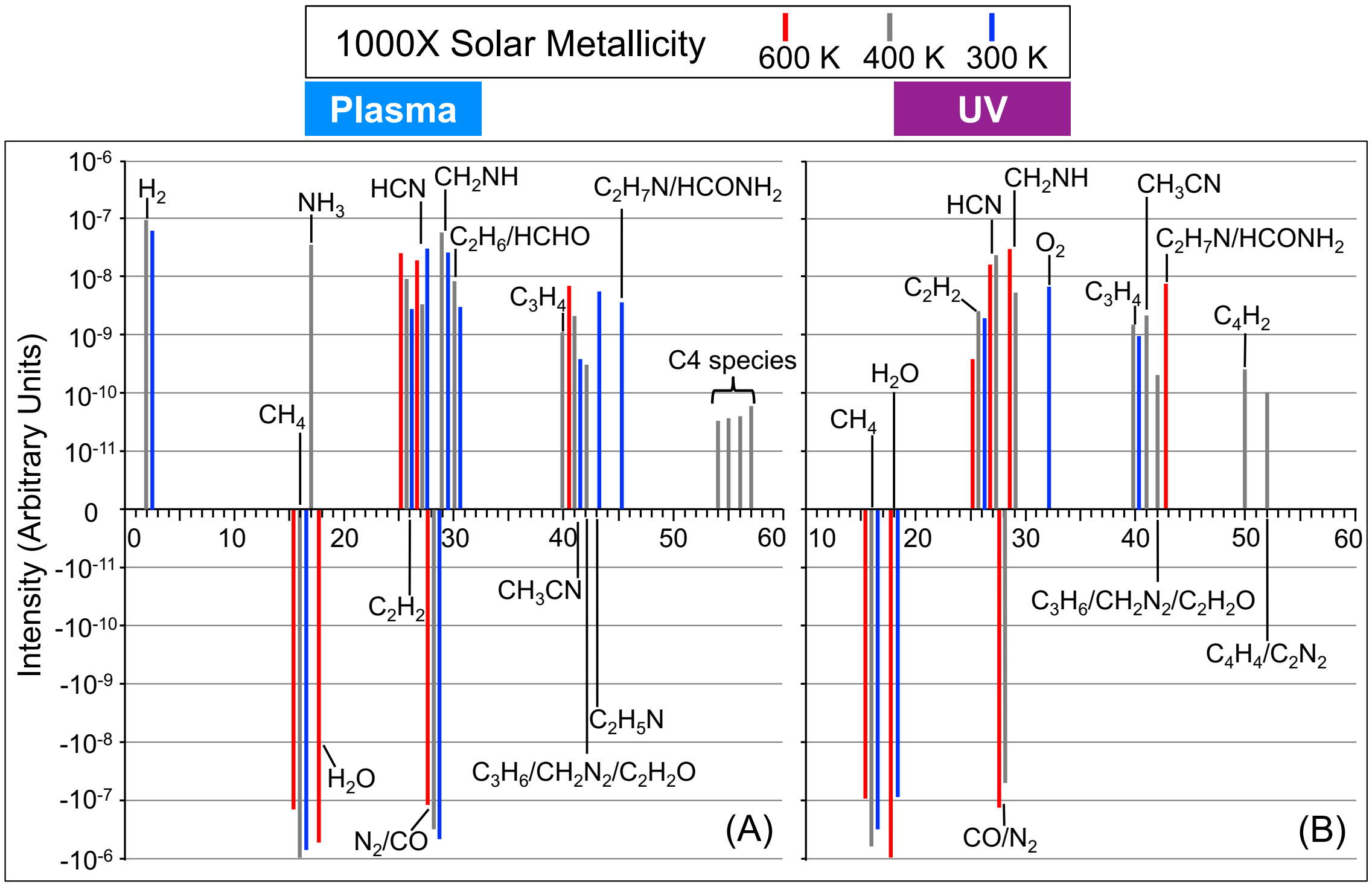

Figure 7. The changed gas peaks in the 1000× metallicity experiments for both energy sources: plasma (A) and UV (B). Results are shown in different colors for different temperatures: red, 600 K; orange, 400 K; blue, 300 K. The chemical formulas are labeled near the peaks (Peaks are offset from the nominal mass to show three spectra in one plot).

The starting gas mixtures for the 1000× metallicity experiments at 400 K and 300 K are $H_2O$-dominated, and are made up of the same gases. However, the mixing ratios of the gases in the mixtures are different, which could drastically affect the gas phase chemistry. As demonstrated in previous studies, the concentration of $CH_4$ or CO in the initial gas mixtures has a dramatic impact on the gas phase chemistry and the solid phase compositions.[17,21] As we discussed above (Section 3.1), in the 400 K-1000× experiment with plasma, we observed decreases for $CH_4$ and $N_2$ in the initial gas mixture, and increases for $H_2$ and newly formed species, including $NH_3$, C2 ($C_2H_2$, HCN, $CH_2NH$, $C_2H_6$/HCHO), C3 [$C_3H_4$ (40), $CH_3CN$ (41), $C_3H_6$/$CH_2N_2$/$C_2H_2O$ (42)], and C4 [$C_4H_6$/$C_2H_2N_2$/$C_3H_2O$ (54); $C_3H_5N$/$C_2HON$ (55); $C_4H_8$/$C_2H_4N_2$/$C_3H_4O$ (56); $C_3H_7N$/$CH_3N_3$/$C_2H_3ON$ (57)] species. In the 300 K-1000× experiment with plasma, we also noticed decreases for $CH_4$ and $N_2$, and increases for $H_2$, C2, and C3 species. The C2 ($C_2H_2$, HCN, $CH_2NH$, $C_2H_6$/HCHO) species are the same as those in the 400 K experiment, but the C3 species are different [$C_2H_5N$, $C_2H_7N$/$HCONH_2$ (45) besides $CH_3CN$]. In the 400 K-1000× experiment with UV, $CH_4$ and $N_2$ in the initial gas mixture decrease, while new C2 ($C_2H_2$, HCN, $CH_2NH$) and C3 ($C_3H_4$, $CH_3CN$, $C_3H_6$/$CH_2N_2$/$C_2H_2O$) species are produced in the gas phase. In the 300 K-1000× experiment with UV, $CH_4$ and $H_2O$ decrease, while $C_2H_2$, $O_2$, and $C_3H_4$ are formed. Again, $C_2H_2$ is formed in all of the 1000× metallicity experiments. HCN is formed in most of these experiments except the UV experiment at 300 K. We noticed that $O_2$ was only produced in the 300 K-1000× UV experiment, but not in the 400 K-1000× UV experiment, although both experiments started with $H_2O$-rich gas mixtures. The photodissociation of water could be a major source of $O_2$. There is less $CH_4$ in the 300 K experiment (6.6%) than in the 400 K experiment (11%), which could result in more UV photons in 115 to 140 nm (Figure 3) remaining to dissociate $H_2O$ and produce $O_2$. More detailed study is required to fully understand the possible mechanisms.

As we did for the 100× metallicity experiments, we show the end products of the 1000× metallicity experiments in Figure 8. For the 1000× metallicity experiments, the initial gases at 600 K are different from those at 400 K and 300 K ($H_2$-dominated versus $H_2O$-dominated). At 600 K, both $CH_4$ and CO can serve as carbon source, while $H_2O$ and $N_2$ are the oxygen and nitrogen sources. At 400 K and 300 K, ample $H_2O$ molecules provide oxygen source, while the initial $CH_4$ and $N_2$ serve as carbon and nitrogen sources. Although we did not observe significant decrease of $CO_2$ in the 1000× metallicity experiments, $CO_2$ could serve as an additional source for carbon and oxygen in these experiments. For the UV experiment at 300 K, we did not observe the decrease of $N_2$ or the formation of any nitrogen-containing molecule in the gas phase, which could result in no nitrogen in the solid phase ($C_xH_yO_z$). The haze production rates (~10 mg/hr) of the 1000× metallicity experiments at 400 K and 300 K are the highest among all the experiments in our study, even higher than that of our standard Titan experiment (~7.4 mg/hr, 5% $CH_4$ in 95% $N_2$) using the PHAZER chamber[17]. This indicates that $H_2O$-dominated gas mixtures efficiently produce haze particles with cold plasma as an energy source. These two gas mixtures also have a higher $CH_4$ content than our standard Titan experiment. The photolysis of $H_2O$ and/or $CH_4$ can produce $H_2$. The increase/production of $H_2$ in the gas phase indicates that $H_2O$ and/or $CH_4$ could be important contributors for generating big organic molecules in our 1000× metallicity plasma experiments at 400 K and 300 K. The production rate is not simply a function of $CH_4$ content. The nitrogen-containing molecules identified in the gas phase (Figure 7), such as $NH_3$, HCN, $CH_2NH$, could also play an important role in haze particle formation in our current experiments, since previous studies have shown that these species are important precursors for Titan's organic aerosol analogues[29-32]. Although the UV lamp (110–400 nm) cannot directly dissociate $N_2$, the nitrogen-containing molecules are produced in the UV experiments. The nitrogen could participate in the photochemistry processes through the reaction of $N_2$ with CH (ground or excited state) formed from $CH_4$ dissociation. A detailed discussion of possible mechanisms can be found in previous studies[19,21].

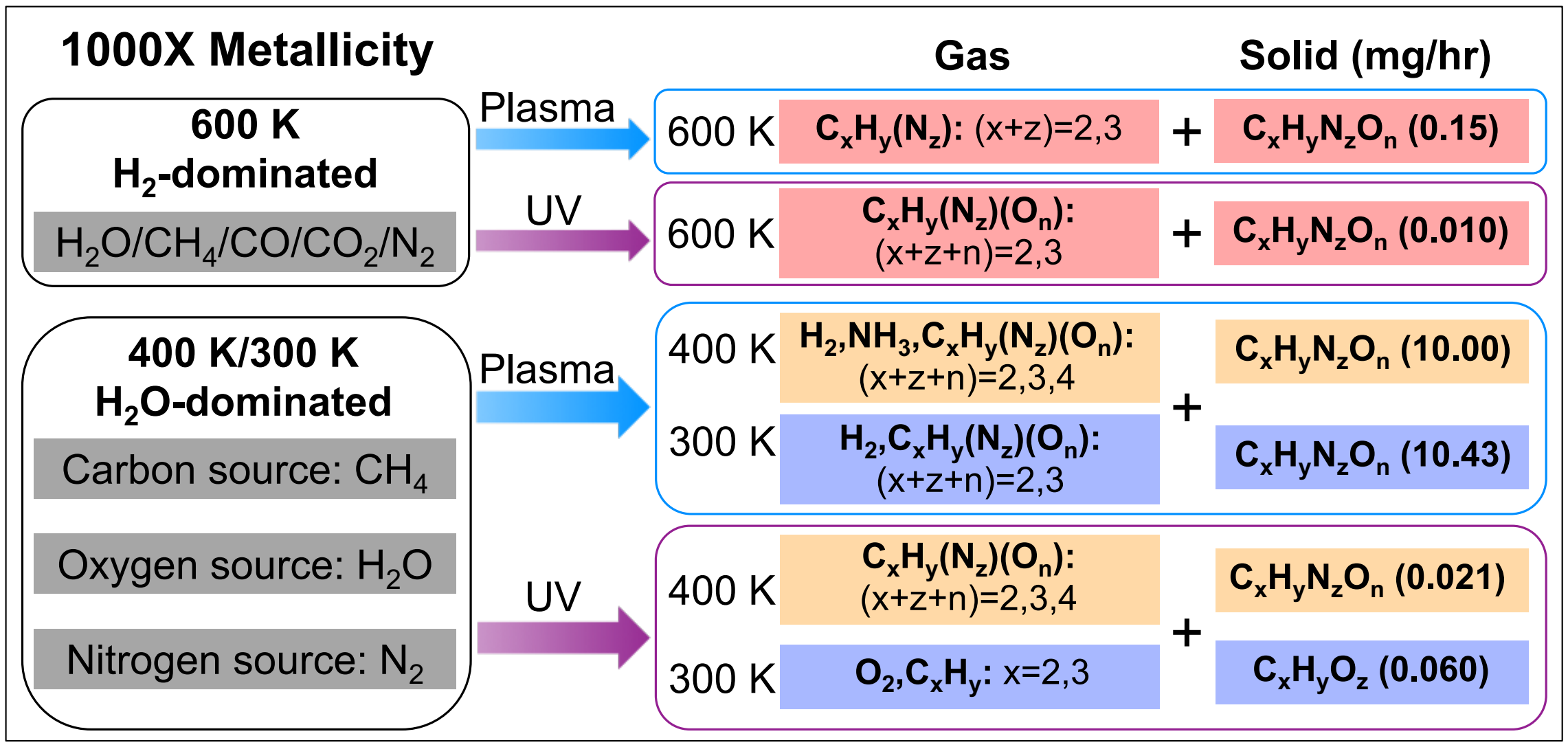

Figure 8. End products of the 1000× metallicity experiments. The chemical formula with parentheses represents multiple forms; for example, $C_xH_y(N_z)(O_n)$ includes $C_xH_y$, $C_xH_yN_z$, $C_xH_yO_n$, and $C_xH_yN_zO_n$. The number in the right column is the haze production rate for each experiment.

Miller[33] showed that the presence of $H_2O$ (as liquid and gas) could promote the formation of organics. As shown in Figure 7, mass peaks 30 and 42 could be contributed or partially contributed to by HCHO (formaldehyde) and $C_2H_2O$ (ketene, $H_2C=C=O$). These oxygen-containing molecules can be produced from the photochemistry of $H_2O$ and/or $CO_2$ in the gas phase and can react with a wide variety of organic molecules and convert to other compounds; therefore, they could play a role in the formation of large organic molecules and haze particles. The difference in gas products formed at 400 K and 300 K (Figure 7) suggests that the chemical pathways and resulting solid products may be also distinct. More detailed work is required to understand the complex chemical processes in the gas mixtures that yield so many haze particles.

*3.4. Gas Molecule Changes during the 10000× metallicity Experiments*

Table 4. The initial gas compositions for the 10000× solar metallicity experiments.

| 600 K | 400 K | 300 K |
|---|---|---|
| 66% $CO_2$ | 67% $CO_2$ | 67.3% $CO_2$ |
| 12% $N_2$ | 15% $H_2O$ | 15.6% $H_2O$ |
| 8.6% $H_2$ | 13% $N_2$ | 13% $N_2$ |
| 5.9% $H_2O$ | 5% He | 4.1% He |
| 3.4% CO | | |
| 4.1% He | | |

The initial gas compositions for the 10000× metallicity experiments are shown in Table 4. Figure 9 shows the changed gas peaks in the 10000× metallicity experiments that started with $CO_2$-dominated gas mixtures. For the 600 K experiment with plasma, $H_2$, $H_2O$, and $CO/N_2$ in the initial gas mixtures decrease, while HCN, $C_2H_6$/HCHO, $O_2$ and $CH_3CN$ increase. For the 600 K experiment with UV, $H_2$ and $CO/N_2$ also decrease, but $H_2O$ increases along with newly formed $C_2H_6$/HCHO, $O_2$ and $C_3H_6/CH_2N_2/C_2H_2O$. For the 400 K and 300 K experiments with plasma, we observed a decrease of $H_2O$ and increases of $C_2H_6$/HCHO and $O_2$. For the 400 K UV experiment, we only saw $H_2O$ decreasing and $O_2$ increasing. At 300 K, besides the change of $H_2O$ and $O_2$, we also observed increases of $C_2H_6$/HCHO and $C_2H_6O$/HCOOH (formic acid). Bar-Nun & Chang[34] also reported that $C_2H_6$ and HCHO were produced from irradiation of CO and $H_2O$. Since there is **neither** $CH_4$ in the initial gas mixtures nor detectable amounts of $CH_4$ being produced in the gas phase, the newly formed C2 and C3 species are more likely to be O-containing organic molecules rather than pure hydrocarbons. For example, m/z 30 is probably from HCHO and m/z 42 is probably from $C_2H_2O$ (ketene, $H_2C{=}C{=}O$). HCHO can be produced through photochemical reactions of CO and $H_2O$[34], or by photochemical reduction of $CO_2$ with $H_2O$[35,36] via net Reaction 1 ($CO_2 + H_2O \Leftrightarrow HCHO + O_2$). HCHO can further react to form a variety of organic compounds, including alcohols, aldehydes, acetone, acetic acid, and more complex organics[34].

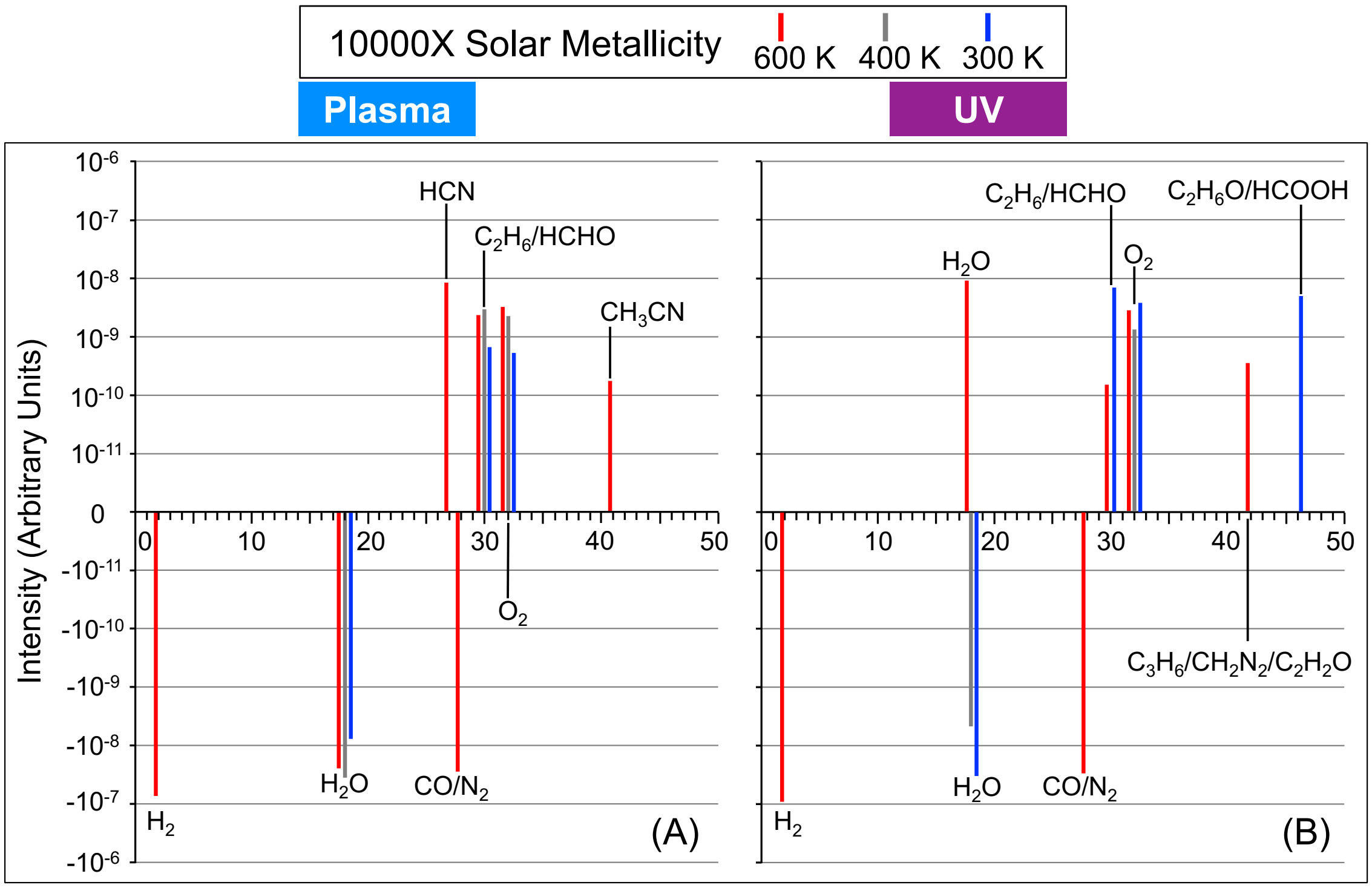

Figure 9. The changed gas peaks in the 10000× metallicity experiments for both energy sources: plasma (A) and UV (B). Results are shown in different colors for different temperatures: red, 600 K; orange, 400 K; blue, 300 K. The chemical formulas are labeled near the peaks (Peaks are offset from the nominal mass to show three spectra in one plot).

Figure 10 shows end products of the 10000× metallicity experiments. The initial gas mixtures are $CO_2$-dominated and relatively oxidizing for the 10000× metallicity experiments. There is no $CH_4$ in the initial gas mixtures, but haze particles are produced at all temperatures with both energy sources. $CO_2$ (and CO at 600 K) provides the carbon source in these experiments. The formation of both new gas products and haze particles demonstrates that $CH_4$ is not necessarily required for organic haze formation. Newly formed gas products (HCN, HCHO, $CH_3CN$, and $C_2H_6O$/HCOOH) could serve as the precursors, suggesting that there are possible new pathways for organic haze formation. Specifying the source of carbon ($CH_4$, CO, or $CO_2$) is not sufficient to determine the complex atmospheric chemistry. The total oxidizing/reducing environments, the ratios of C/H/O/N, and the forms of other species are all involved in the chemical processes. The experiment at 600 K with plasma has the highest haze production rate among the 10000× metallicity experiments, although they all have relatively low production rate. The N-containing molecules in the gas phase (HCN, $CH_3CN$) indicate N chemistry may contribute to the higher haze production rate in this case.

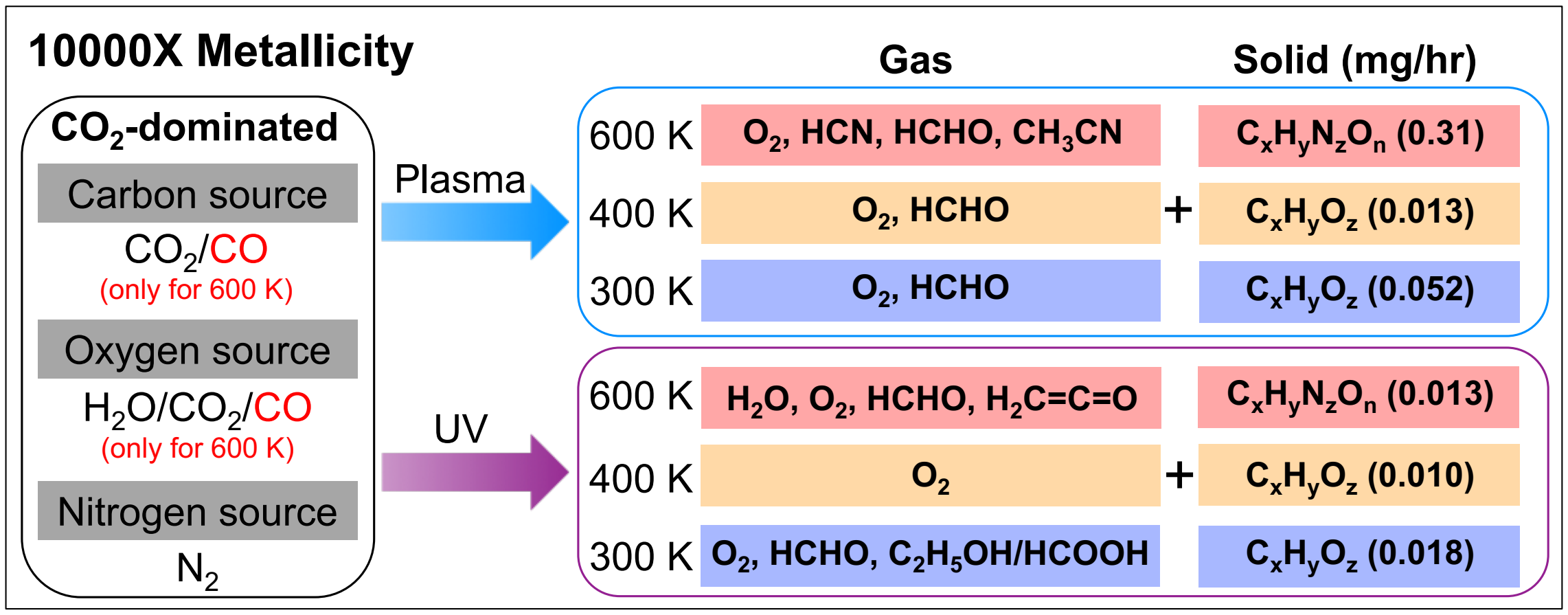

Figure 10. End products of the 10000× metallicity experiments. The chemical formula with parentheses represents multiple forms; for example, $C_xH_y(N_z)(O_n)$ includes $C_xH_y$, $C_xH_yN_z$, $C_xH_yO_n$, and $C_xH_yN_zO_n$. The number in the right column is the haze production rate for each experiment.

### 3.5 Relative yields of gas products

The experiments started with different gas mixtures and generated different gas products. Because these species have different electron-impact ionization cross sections and the mass spectrometer has different instrumental responses for these species, the peaks for different species are not quantitative. Thus, it is very difficult to calculate the absolute yield of gas products from the mass spectra. In order to compare the yields of the gas product between experiments, we estimated the relative yield of gas products by assuming the molecular peaks are quantitative for the gas products (In other words, we assume that the cross sections are the same for all the species, and all species follow the same linear calibration curves on the RGA mass spectrometer). Due to the variety of gas products in different experiments, we calculate the yield by combining the peak intensity increase ($I$) and the number of heavy atoms (non-H atoms) in the gas products ($N$). For each experiment, the total yield of the gas products ($Y$) equals:

$$Y = \sum_{k=0}^{k} I_k \, N_k$$

where $I_k$ is the intensity increase for each gas product, and $N_k$ is the number of heavy atoms (non-H atoms) in the gas product.

We normalized the total yield of gas products for each experiment by using the highest yield experiment as reference (100%) and obtained the relative yields for all 18 experiments (Figure 11). Figure 11 shows that the 400 K-1000× metallicity UV and plasma experiments yield the most and second most gas products, respectively. This could be due to the highest mixing ratio of $CH_4$ in the 400 K-1000× metallicity experiments, but other gases ($N_2$, $CO_2$, $H_2O$) in the initial gas mixture can also have strong influences on the photochemistry. The 300 K-10000× metallicity plasma experiment yields the least gas products out of the 18 experiments (both plasma and UV), which is two orders of magnitude lower than the highest yield case. The 400 K-10000× metallicity UV experiment is the lowest among the nine UV experiments. The low yields of gas products and low production rates of solid particles in the 10000× metallicity experiments may be due to the oxidizing environment of the initial gas mixtures. The only carbon source ($CO_2$) is oxidized in the initial gas mixtures of 10000× metallicity experiments at 400 K and 300 K (Figure 2).

We have reported the haze production rates in these experiments[13,14]. We normalized the haze production rates by using the highest production rate experiment (300 K-1000× metallicity plasma experiment) as reference (100%) and plotted the relative yields of solid products (haze particles) in Figure 11 along with the gas products. There is a positive correlation between the yields of new gas products and the production rates of solid particles for the plasma experiments. The two experiments that have highest yields of gas products also generate the most solid particles. Such correlation is not observed in the UV experiments. Figure 11 shows that the gas phase production yields are on the same order of magnitude for the plasma and UV experiments for most cases. However, in 8 out of 9 cases, the solid particle production rates of the plasma experiments are higher than those of the UV experiments[13,14] with only one exception (the 300 K-100× case). The 300 K-100× case has a better solid phase production yield with UV, which is also the only gas mixture that contains $NH_3$. $NH_3$ could play an important role in the production of solid particles, and we will look into its role in future investigations. Our results indicate that the conversion from gas to solid phase is much more efficient in the plasma experiments than in the UV experiments, for 8 out of 9 cases. The higher haze production rates in the plasma experiments could reflect the higher energy density of the AC glow discharge than that of the UV photons. It should be noted that the experiments started with nine different gas mixtures. The photochemical processes in nine cases with two different energy sources are very complex, and the nucleation and growth of the solid particles are related to temperature and vapor pressure as well. Therefore, more comprehensive study is necessary to understand the complex chemical and physical processes happening in these atmospheres.

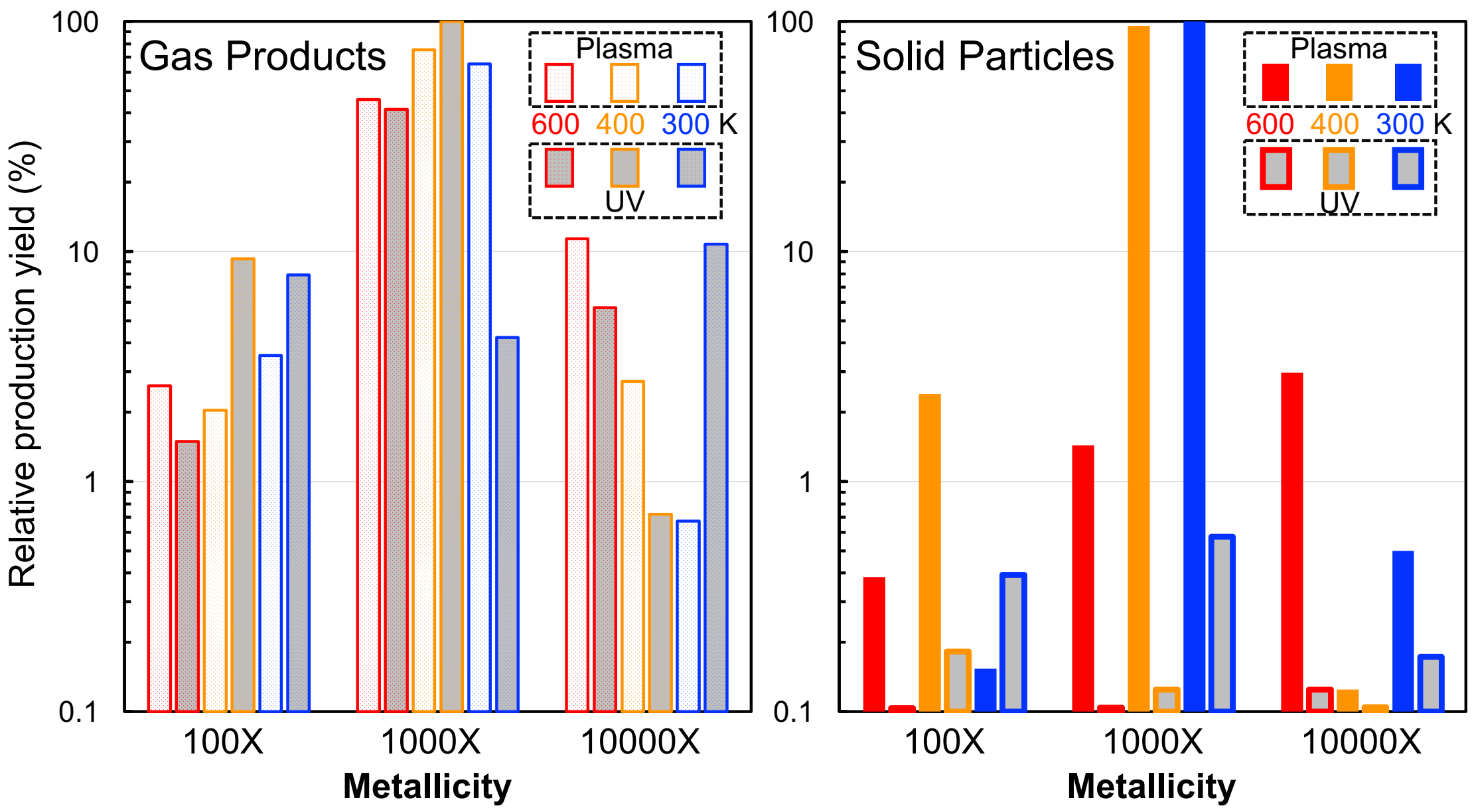

Figure 11. Relative yield of gas products and solid products (haze particles,) in the plasma and UV experiments. The 400 K-1000× metallicity UV experiment has the highest gas products yield and is used as a reference (100%) for gas products yields in other experiments. The 300 K-1000× metallicity plasma experiment has the highest haze production rate and is used as a reference (100%) for solid products (haze) yields in other experiments. Note that three UV experiments (600 K-100×, 600 K-1000×, and 400 K-10000×) have low haze production rate and their relative yields are lower but very close to 0.1%.

### 3.6 Exoplanet Biosignatures

$O_2$ makes up ~20% of Earth's atmosphere, and it is considered one of the most robust biosignature gases in Earth's atmosphere[37] because it would be present only in trace amounts if there is no continual replenishment by photosynthesis by plants and bacteria. Thus, $O_2$ (and its photochemical product ozone, $O_3$) has been considered as a biosignature gas in the search for life beyond our solar system[38]. Models have suggested that photochemically produced $O_2$ can be a potential false positive biosignature in exoplanet atmospheres[39,40]. In our experiments, $O_2$ is formed in all of the 10000× metallicity cases (the initial gas mixtures are most akin to atmospheres expected for terrestrial exoplanets), and also in the 1000× metallicity case at 300 K with UV. The result here clearly demonstrates that $O_2$ can be produced abiotically through photochemical processes in multiple atmospheric scenarios, probably via photochemical reaction of $CO_2$ and $H_2O$ (Reaction 1). Thus, we should reconsider $O_2$ as a biosignature in exoplanet atmospheres; we can assume $O_2$ as a biosignature gas only if we rule out that $O_2$ is generated photochemically or geochemically[40]. However, that is easier said than done because it requires sufficient data to identify atmospheric compositions and constrain a wide variety of possibilities (geochemical settings, surface and atmospheric chemistry, and photochemistry scenarios of exoplanets).

Besides $O_2$, many organic molecules that are produced by life on Earth are also considered to be potential exoplanet biosignatures[41], since these molecules could be produced and accumulate in exoplanet atmospheres. However, these molecules can be also generated abiotically[42]. Our experiments show that some of those organics (such as $C_2H_2$, HCN, HCHO/$C_2H_6$, $CH_3CN$, and $C_2H_6O$/HCOOH) can be formed photochemically. The simultaneous presence of hydrocarbons and $O_2$ (a redox disequilibrium) has been suggested as a sign for life on rocky planets[43.44]. However, we observed the coexistence of $O_2$ and simple organics (could be pure hydrocarbons and/or reduced nitrogen-/oxygen-containing organics) in several simulated atmospheres, which indicates that even the simultaneous presence of $O_2$ and organics could be false positive biosignature. Thus, we must consider the possibility of false positives of these potential biosignature gases. The surface and atmospheric chemistry on exoplanets should be carefully examined and all possible abiotic sources of these gases should be evaluated in those alternative planetary environments.

3.7 Key precursors in the gas phase

As we mentioned above, several newly formed molecules in the gas phase could be very important precursors of larger molecules in the gas phase and solid phase. These species include $C_2H_2$, HCN, $NH_3$, $CH_2NH$, HCHO, and $H_2C{=}C{=}O$. It should be noted that we could not definitively identify HCHO and $H_2C{=}C{=}O$ from only the mass spectra. However, these two species are probably formed in the 10000× metallicity experiments and play a major role for producing haze particles in these cases. Photolysis of $CH_4$ produces $C_2H_2$ that is a key molecule for the production of heavy hydrocarbons, as demonstrated in laboratory simulations of Titan's atmosphere[26,45]. In fact, we observed the formation of $C_2H_2$ in all our experiments in which the gas mixture includes $CH_4$. $C_2H_2$ could be responsible for the formation of heavy hydrocarbons in exoplanet atmospheres in a similar manner as on Titan and giant planets.

The nitrogen-bearing molecules, HCN, $NH_3$, and $CH_2NH$, can be produced from gas mixtures that include $N_2$, and are important for nitrogen chemistry in planetary atmospheres[46]. We detected nitrogen-bearing molecules in our experiments that have $N_2$ and $CH_4$ in the gas mixture. Their photochemical formation pathways may be similar to that in Titan's atmosphere[46]. The polymerization, co-polymerization and/or incorporation of these species could take place in the upper atmosphere of planets, leading to the formation of nitrogenated gas molecules and haze particles[29-32].

The oxygen-bearing species (HCHO and $H_2C{=}C{=}O$) could be derived from $H_2O$, CO, and/or $CO_2$ in the initial gas mixture. HCHO could be an important precursor, which is very reactive under these conditions and can form a variety of organic compounds[34]. Photochemical production of HCHO from atmospheric $CH_4$, CO, and/or $CO_2$ with different energy sources has been reported in previous experiments[25,33,35,47]. In the presence of a dissociating or ionizing energy source, all these potential gas precursors and their photochemical products can react with each other and further react with the resulting products, eventually resulting in particle formation. Efforts are underway to measure composition of the resulted haze particles in order to better understand the complex photochemical processes.

Although observations with current facilities such as the Hubble Space Telescope (the observations by Hubble were wavelength-limited, 0.5 to 1.7 microns) have largely revealed featureless transmission spectra for mini-Neptunes and super-Earths, future ground-based facilities and JWST will be able to probe spectral features of major gas compositions and possible clouds/hazes in exoplanet atmospheres. Gas phase species composition is often easier to measure using remote sensing techniques than solid phase species, particularly for complex solids like the photochemical hazes formed in our experiments. When we can only measure gas phase species in an exoplanet atmosphere, the detection of those species may still provide information about composition of particles present in the atmosphere. The gas precursors we identified here could be important atmospheric chemical indictors of photochemistry and haze formation in exoplanet atmospheres.

In addition, HCHO and HCN are also important prebiotic precursors since they readily undergo a variety of reactions to give products of biological significance including sugars, amino acids, and nucleobases[36,48-52]. Since both HCHO and HCN are formed in several of our experiments, we will particularly search for prebiotic molecules (sugars, amino acids, and nucleobases) during our composition analysis of the resulting solid materials.

## 4. CONCLUSIONS

We conducted laboratory experiments with the PHAZER chamber to simulate haze formation in cool exoplanet atmospheres and monitored the gas phase chemistry happening during these experiments. Using mass spectroscopy, we measured a variety of new gas phase species formed from diverse initial gas mixtures that we subjected to either energetic cold plasma or UV photons. We found that the yields of the gas products vary greatly in the experiments with different initial gas mixtures, and the energy sources (plasma or UV) clearly affect the chemical processes and the gas product yields in the experiments starting with the same gas mixture. There is a positive correlation between the yields of new gas products and the production rates of solid particles for the plasma experiments, but such correlation is not observed in the UV experiments.

Combining the gas composition change and the haze production rate in each experiment,

we suggested possible chemical processes that lead to the formation of haze particles, which were distinct for the experiments with different initial gas mixtures. There are multiple chemical pathways to haze formation. Additionally, we observed new organic gas molecules and haze formation in the 10000× metallicity experiments that do not have $CH_4$ in the initial gas mixtures, confirming that $CH_4$ is not necessarily required for the formation of organics (gas and/or solid phase) and $CO_2$ (and CO at 600 K) can provide an alternative source of carbon for organics.

We identified some key gas precursors ($C_2H_2$, HCN, $NH_3$, $CH_2NH$, HCHO, and $H_2C{=}C{=}O$) that are indicative of haze formation and could be detected by JWST. $O_2$ and organic gas products are formed through photochemical processes in simulated atmospheric scenarios, demonstrating that we should eliminate abiotic production channels of these molecules before considering them as biosignatures in exoplanet atmospheres. In conclusion, our results show that complex atmospheric photochemistry can happen in diverse exoplanet atmospheres, and produce new gas products and haze particles. The gas phase products we identified here are very important to understand the gas-solid composition connections and the possible chemical processes that lead to the haze formation in exoplanet atmospheres.

## 5. ACKNOWLEDGEMENTS

This work was supported by the NASA Exoplanets Research Program Grant NNX16AB45G. C.H. was supported by the Morton K. and Jane Blaustein Foundation. J.M. acknowledges support from NASA grant NNX16AC64G.

## REFERENCES

1. Elkins-Tanton, L. T.; Seager, S. Ranges of Atmospheric Mass and Composition of Super-Earth Exoplanets. *Astrophys. J.* **2008**, *685,* 1237–1246.
2. Schaefer, L.; Lodders, K.; Fegley, B. JR. Vaporization of the Earth: Application to Exoplanet Atmospheres. *Astrophys. J.* **2012**, *755,* 41.
3. Moses, J. I.; Line, M. R.; Visscher, C.; et al. Compositional diversity in the atmospheres of hot Neptunes, with application to GJ 436b. *Astrophys. J.* **2013**, *777,* 34.

4. Hu, R; Seager, S. Photochemistry in Terrestrial Exoplanet Atmospheres III: Photochemistry and Thermochemistry in Thick Atmospheres on Super Earths. *Astrophys. J.* **2014**, *784,* 63.
5. Ito, Y.; Ikoma, M.; Kawahara, H.; Nagahara, H.; Kawashima, Y.; Nakamoto, T. Theoretical Emission Spectra of Atmospheres of Hot Rocky Super-Earths. *Astrophys. J.* **2015**, *801,* 144.
6. Kreidberg, L., Bean, J. L., Désert, J.-M., et al. Clouds in the atmosphere of the super-Earth exoplanet GJ1214b. *Nature* **2014**, 505, 69–72.
7. Knutson, H. A.; Benneke, B.; Deming, D.; Homeier, D. A featureless transmission spectrum for the Neptune-mass exoplanet GJ 436b. *Nature* **2014**, 505, 66–68.
8. Knutson, H. A.; Dragomir, D.; Kreidberg, L.; et al. Hubble Space Telescope Near-IR Transmission Spectroscopy of the Super-Earth HD 97658b. *Astrophys. J.* **2014**, *794,* 155.
9. Lothringer, J. D.; Benneke, B.; Crossfield, I. J. M.; et al. An HST/STIS Optical Transmission Spectrum of Warm Neptune GJ 436b. *Astron. J.* **2018,** 155, 66
10. Dragomir, D.; Benneke, B.; Pearson, K. A.; et al. Rayleigh Scattering in the Atmosphere of the Warm Exo-Neptune GJ 3470b. *Astrophys. J.* **2015**, *814,* 102.
11. Marley, M. S.; Ackerman, A. S.; Cuzzi, J. N.; Kitzmann, D. in *Comparative Climatology of Terrestrial Planets*, ed. S. J. Mackwell et al. **2013**, (Tucson, AZ: Univ. Arizona Press), 367
12. Morley, C. V.; Fortney, J. J.; Marley, M. S.; et al. Thermal Emission and Albedo Spectra of Super Earths with Flat Transmission Spectra. *Astrophys. J.* **2015**, *815,* 110.
13. He, C.; Hörst, S. M.; Lewis, N. K.; et al. Photochemical Haze Formation in the Atmospheres of super-Earths and mini-Neptunes. *Astron. J.* **2018,** 156, 38.
14. Hörst, S. M.; He, C.; Lewis, N. K.; et al. Haze production rates in super-Earth and mini-Neptune atmosphere experiments. *Nature Astronomy* **2018,** *2*, 303–306.
15. Cable, M. L.; Hörst, S. M.; Hodyss, R.; et al. Titan tholins: simulating Titan organic chemistry in the Cassini-Huygens era. *Chem. Rev.* **2012,** *112,* 1882–1909.
16. He, C.; Hörst, S. M.; Lewis, N. K.; et al. Laboratory Simulations of Haze Formation in the Atmospheres of Super-Earths and Mini-Neptunes: Particle Color and Size Distribution. *Astrophys. J. Lett.* **2018,** *856,* L3.
17. He, C.; Hörst, S. M.; Riemer, S.; Sebree, J. A.; Pauley, N.; Vuitton, V. Carbon Monoxide Affecting Planetary Atmospheric Chemistry. *Astrophys. J. Lett.* **2017,** *841,* L3.
18. Trainer, M. G.; Pavlov, A. A.; DeWitt, H. L.; et al. Organic haze on Titan and the early Earth. *Proc. Natl. Acad. Sci. U. S. A.* **2006,** *103,* 18035–18042.
19. Trainer, M. G.; Jimenez, J. L.; Yung, Y. L.; Toon, O. B.; and Tolbert, M. A. Nitrogen Incorporation in $CH_4$-$N_2$ Photochemical Aerosol Produced by Far Ultraviolet Irradiation. *Astrobiology* **2012,** *12,* 315–326.
20. Sebree, J. A., Trainer, M. G., Loeffler, M. J., Anderson, C. M. Titan aerosol analog absorption features produced from aromatics in the far infrared. *Icarus* **2014,** *236*, 146–152.
21. Hörst, S. M.; Yoon, Y. H.; Ugelow, M. S.; et al. Laboratory Investigations of Titan Haze Formation: In Situ Measurement of Gas and Particle Composition. *Icarus* **2018,** *301*, 136–151.
22. Hörst, S. M.; He, C.; Ugelow, M. S.; et al. Exploring the Atmosphere of Neoproterozoic Earth: The Effect of $O_2$ on Haze Formation and Composition. *Astrophys. J.* **2018**, *858,* 11.
23. Hodyss, R.; Howard, H. R.; Johnson, P. V.; Goguen, J. D.; Kanik, I. Formation of radical

species in photolyzed $CH_4$:$N_2$ ices. *Icarus* **2011,** *214*, 748–753.
24. DeWitt, H. L.; Trainer, M. G.; Pavlov, A. A.; Hasenkopf, C. A.; Aiken, A. C.; Jimenez, J. L.; McKay, C. P.; Toon, O. B.; Tolbert, M. A. Reduction in haze formation rate on prebiotic Earth in the presence of hydrogen. *Astrobiology* **2009,** *9,* 447–453.
25. Chyba, C.; Sagan, C. Endogenous production, exogenous delivery and impact- shock synthesis of organic molecules: an inventory for the origins of life. *Nature* **1992,** *355,* 125–132.
26. Coll, P.; Coscia, D.; Smith, N.; Gazeau, M.; Ramírez, S. I.; Cernogora, G.; Israël, G.; Raulin, F. *Planet. Space Sci.* **1999,** *47*, 1331–1340.
27. Yung, Y. L.; Allen, M.; Pinto, J. P. Photochemistry of the atmosphere of Titan: comparison between model and observations. *Astrophys. J., Suppl. Ser.* **1984,** *55,* 465–506.
28. Wilson, E. H.; Atreya, S. K.; Coustenis, A. Mechanisms for the formation of benzene in the atmosphere of Titan. 2003, *J. Geophys. Res.: Planets* **2003**, *108*, 5014.
29. He, C.; Lin, G.; Upton, K.T.; Imanaka, H.; Smith, M. A. Structural Investigation of Titan Tholins by Solution-State $^{1}H$, $^{13}C$, and $^{15}N$ NMR: One-Dimensional and Decoupling Experiments. *J. Phys. Chem. A* **2012,** *116*, 4760–4767.
30. He, C.; Smith, M.A. Identification of nitrogenous organic species in Titan aerosols analogs: Nitrogen fixation routes in early atmospheres. *Icarus* **2013,** *226,* 33–40.
31. He, C. & Smith, M.A. A Comprehensive NMR Structural Study of Titan Aerosol Analogs: Implications for Titan's Atmospheric Chemistry. *Icarus* **2014,** *238,* 31–38.
32. He, C. & Smith, M.A. Identification of nitrogenous organic species in Titan aerosols analogs: Implication for Prebiotic Chemistry on Titan and Early Earth. *Icarus* **2014,** *238,* 86–92.
33. Miller, S. L. A production of amino acids under possible primitive Earth conditions. *Science* **1953,** *117,* 528–529.
34. Bar-Nun, A; Chang, S. Photochemical reactions of water and carbon monoxide in Earth's primitive atmosphere. *J. Geophys. Res. Oceans* **1983,** *88,* 6662–6672.
35. Pinto, J. P.; G. R. Gladstone; Y. L. Yung. Photochemical production of formaldehyde in earth's primitive atmosphere. *Science,* **1980**, *210,* 183–185.
36. Cleaves, H. J. The prebiotic geochemistry of formaldehyde. *Precambrian Research* **2008,** *164,* 111–118.
37. Léger, A.; Pirre, M.; Marceau, F. J. Search for primitive life on a distant planet: Relevance of $O_2$ and $O_3$ detections. *Astron. Astrophys.* **1993,** *277,* 309–313.
38. Seager, S.; Bains, W. The search for signs of life on exoplanets at the interface of chemistry and planetary science. *Sci. Adv.* **2015,** *1,* e1500047.
39. Hu, R.; Seager, S.; Bains, W. Photochemistry in Terrestrial Exoplanet Atmospheres I: Photochemistry Model and Benchmark Cases, *Astrophys. J.* **2012,** *761,* 166
40. Domagal-Goldman, S. D.; A. Segura, M. W. Claire, T. D. Robinson, Meadows V. S. Abiotic ozone and oxygen in atmospheres similar to prebiotic Earth. *Astrophys. J.* **2014,** *792,* 90–104.
41. Seager, S.; Bains, W.; Petkowski, J. J. Toward a List of Molecules as Potential Biosignature Gases for the Search for Life on Exoplanets and Applications to Terrestrial Biochemistry. *Astrobiology* **2016,** *16,* 465–485.
42. Schwieterman, E. W.; Kiang, N. Y.; Parenteau, M. N.; et al. Exoplanet Biosignatures: A Review of Remotely Detectable Signs of Life. *Astrobiology* **2018,** *18,* 663–708.
43. Lippincott E. R.; R. V. Eck, M. O. Dayhoff; Sagan, C. Thermodynamic equilibria in

planetary atmospheres. *Astrophys. J.* **1967,** *147,* 753–764.
44. Lovelock J. E.; Kaplan, I. R. Thermodynamics and the recognition of alien biospheres. *Proc. Roy. Soc. Lond. B.* **1975,** *189,* 167–181.
45. Cabane, M.; Chassefière, E. Laboratory simulations of Titan's atmosphere: organic gases and aerosols. *Planet. Space Sci.* **1995,** *43*, 47–65
46. Vuitton, V.; Yelle, R. V.; Anicich, V. G. The Nitrogen Chemistry of Titan's Upper Atmosphere Revealed. *Astrophys. J.* **2006,** *647,* L175– L178
47. Stribling, R.; Miller, S. Energy yields for hydrogen cyanide and formaldehyde syntheses: the hydrogen cyanide and amino acid concentrations in the primitive ocean. *Orig. Life Evol. Biosph.* **1987,** *17,* 261–273.
48. Miller, S. L. The mechanism of synthesis of amino acids by electric discharges. *Biochim Biophys Acta* **1957,** *23,* 480–489.
49. Schwartz, A. W.; Voet, A. B.; Van Der Veen, M. Recent progress in the prebiotic chemistry of HCN. *Origins of Life* **1984,** *14,* 91–98.
50. Leslie, O. E. Prebiotic Chemistry and the Origin of the RNA World, *Crit. Rev. Biochem. Mol. Biol.* **2004,** *39,* 99–123.
51. Feng, S.; Tian, G.; He, C.; et al. Hydrothermal biochemistry: from formaldehyde to oligopeptides. *J. Mater. Sci.* **2008,** *43,* 2418–2425.
52. Hörst, S. M.; Yelle, R. V.; Buch, A.; et al. Formation of amino acids and nucleotide bases in a Titan atmosphere simulation experiment. *Astrobiology* **2012,** *12,* 809–817.